\documentclass[aps,prd,twocolumn,superscriptaddress,nofootinbib]{revtex4-1}

\usepackage{latexsym}
\usepackage{amsmath}
\usepackage{amssymb}
\usepackage{amsfonts}
\usepackage{bm}
\usepackage{physics}

\usepackage{color}
\definecolor{purple}{rgb}{0.5,0,0.5}
\definecolor{blue}{rgb}{0.0,0,0.9}
\definecolor{prdblue}{rgb}{0.133,0.118,0.498}
\usepackage[colorlinks=true, pdfstartview=FitV, linkcolor=prdblue, citecolor= prdblue, urlcolor=prdblue]{hyperref}

\usepackage{supertabular} 
\usepackage{placeins}
\usepackage{epsfig}
\usepackage{graphicx}

\begin{document}


\title{Heavy multiquark systems as clusters of smaller units \\ -- a diffusion Monte Carlo calculation --}


\author{M.C. Gordillo}
\email[]{cgorbar@upo.es}
\affiliation{Departamento de Sistemas F\'isicos, Qu\'imicos y Naturales, Universidad Pablo de Olavide, E-41013 Sevilla, Spain}
\affiliation{Instituto Carlos I de Física Teórica y Computacional,
Universidad de Granada, E-18071 Granada, Spain.}

\author{J. Segovia}
\email[]{jsegovia@upo.es}
\affiliation{Departamento de Sistemas F\'isicos, Qu\'imicos y Naturales, Universidad Pablo de Olavide, E-41013 Sevilla, Spain}


\date{\today}

\begin{abstract}
Multiquark systems appear less frequently than mesons and baryons despite the enormous world-wide experimental effort that has been made during the last two decades. In this work, we will propose a possible explanation for that fact, restricting ourselves to the case of sets including only $c$ and $\bar{c}$ quarks. We will show that those multiquarks can be thought as different combinations of smaller units that associate together to produce colorless assemblies with a definite value of the total spin. For instance, for the $cccccc$ hexaquark with $S=0$, we have three possibilities: a set of six undistinguishable $c$ quarks, an association of two $ccc$ baryons, or a set of three $cc$ diquarks close together. This means we can have three different values for the mass of an open-charm hexaquark with $S=0$. Using the diffusion Monte Carlo method, we calculate all possible combinations compatible with tetraquark $cc \bar{c} \bar{c}$, pentaquark $cccc \bar{c}$, open-charm $cccccc$ and hidden-charm  $ccc \bar{c} \bar{c} \bar{c}$ hexaquark structures with the minimum value of total spin ($S=0$ or $S=1/2$). We consider compact structures with radial wave functions including interactions between all the quarks in the cluster. We find that, in all cases, the mass of the multiquark decreases with the number of small units that conform the set of quarks. For instance, an open charm hexaquark made up of three diquarks has a smaller mass than a set of six of $c$ undistinguishable units. When the pieces that conform the multiquark are themselves colorless with a definite value of the total spin, the cluster splits into those smaller units that separate infinitely from each other.  
\end{abstract}


\maketitle


\noindent\emph{1.- Introduction}.\,---\, A century of fundamental research in atomic physics has demonstrated that ordinary matter is corpuscular, with the atoms themselves containing a dense nuclear core composed of protons and neutrons, collectively named as nucleons, which are members of a broader class of femtometre-scale particles, called hadrons. In working towards an understanding of hadrons, it has been discovered that they are bound-states of quarks and gluons whose strong nuclear interactions are described by a Poincar\'e invariant quantum non-Abelian gauge field theory; namely, Quantum Chromodynamics (QCD).

Solving QCD exhibits a fundamental problem, never before have we been confronted by a theory whose elementary color excitations (quarks and gluons) are not those degrees-of-freedom readily accessible via experiment, \emph{i.e.} they always appear confined inside colorless systems (hadrons). This complexity makes hadron spectroscopy, the collection of readily accessible states constituted from quarks and gluons, the starting point for all further investigations. A very successful classification scheme for hadrons in terms of their valence quarks and antiquarks was independently proposed by Murray Gell-Mann~\cite{GellMann:1964nj} and George Zweig~\cite{Zweig:1964CERN} in 1964. It basically separates hadrons in mesons and baryons which are, respectively, quark-antiquark and three-quark bound-states located at the multiplets of the flavor symmetry. The so-called quark model classification received experimental verification in the late 1960s and, despite extensive experimental searches, no unambiguous candidates for other exotic quark-gluon configurations were identified until the beginning of the third millennium.

The Belle collaboration reported in 2003~\cite{Belle:2003nnu} an anomalous signal, named $X(3872)$, in the invariant mass spectrum of $\pi^+ \pi^- J/\psi$ produced in $B^\pm\to K^\pm X(3872)\to K^\pm (\pi^+ \pi^- J/\psi)$ decays. The $X(3872)$ was later studied by the CDF, D0, and BaBar collaborations  confirming that its quantum numbers, mass and decay patterns make it an  unlikely conventional charm--anti-charm (charmonium) candidate. Therefore, the  simplest quark model picture that had been so successful for around $40$ years was challenged leading to an explosion of related experimental and theoretical activity since then. Nowadays, the number of exotic so-called XYZ states has increased dramatically, in both light- and heavy-quark sectors but also with respect to the meson and baryon categories. For extensive recent presentations about the status of exotic hadrons, the reader is referred to several reviews~\cite{Lebed:2016hpi, Chen:2016qju, Chen:2016spr, Ali:2017jda, Guo:2017jvc, Olsen:2017bmm, Liu:2019zoy, Brambilla:2019esw, Yang:2020atz, Dong:2020hxe, Dong:2021bvy, Chen:2021erj, Cao:2023rhu, Mai:2022eur, Meng:2022ozq, Chen:2022asf, Guo:2022kdi, Ortega:2020tng, Huang:2023jec, Lebed:2023vnd, Zou:2021sha}.

The ultimate aim of theory is to describe the properties of the XYZ states from QCD's first principles. Since the strong coupling constant becomes large in the energy regime where hadrons live, perturbative methods are of limited use in QCD. The two most promising \emph{ab-initio} approaches are effective field theories~\cite{Vogl:1991qt, Brambilla:2004jw, Pineda:2011dg, Albaladejo:2012te} and lattice gauge theories~\cite{Dudek:2010wm, Edwards:2011jj, Ryan:2020iog, USQCD:2022mmc}. In fact, they have recently played a major role in reproducing the observed mass spectrum of stable, long-lived, conventional hadrons but, unfortunately, they have also appeared as very limited methods when treating excitations, states close to hadron-hadron thresholds and multiquark structures~\cite{Brambilla:2018pyn, Brambilla:2019jfi, Briceno:2015rlt, Bulava:2024hec}. Therefore, a more modest goal is the development of QCD motivated phenomenological models that specify the colored constituents, how they are clustered and the forces between them. In that line, simultaneously to the experimental measurements, theorists have been proposing for the XYZ states different kinds of color-singlet clusters, made by quarks and gluons, which go beyond conventional mesons and baryons such as glueballs, quark-gluon hybrids and multiquark systems (for a graphic picture of these kinds of hadrons see, for example, Figs.~1, 6, and 7 of Ref.~\cite{Olsen:2017bmm}).

Concerning the multiquark systems, the very first quark model proposals already speculated with their existence~\cite{Jaffe:1975fd, Jaffe:1976ih, Jaffe:1976ig}. In fact, QCD does not forbid to construct more complex colorless arrangements of valence quarks than mesons and baryons, and provides simple mechanisms to construct multiquark structures. For instance, since a diquark (anti-diquark) in a color antisymmetric (symmetric) combination acts as if it were a single antiquark (quark), (anti-)diquarks could thus become the building blocks of compact tetraquarks, $[(qq)(\bar q\bar q)]$, pentaquarks, $[(qq)(qq)\bar q]$, and even hexaquarks, $[(qq)(qq)(qq)]$, whose size is of the order of the confining scale. A further QCD mechanism for the creation of multiquark structures is inspired by the residual strong interaction that binds nucleons in nuclei, \emph{i.e.} the nuclear binding is effective in meson-meson, baryon-meson and baryon-baryon combinations so as to produce loosely bound and extended molecular-type of multiquark systems. These various quark binding mechanisms could lead to different exotic families, or even to systems with mixed features.

One important observation related with multiquark systems is that they appear much less frequently than usual mesons and baryons despite the enormous world-wide experimental effort that has been made since mid 1960s but specially in the last two decades. The goal of the present theoretical study is to shed some light about this fact performing stability assessments of exotic multiquark structures such as tetraquarks, pentaquarks and hexaquarks assuming all possible clusters between quarks and antiquarks as building blocks. This should yield a hierarchy among the different organizations/families and potentially an explanation of why mesons and baryons have been the only hadron states discovered for decades and are still overwhelming abundant nowadays.

In order to comply with our aim, we use a diffusion Monte Carlo method (DMC) to solve the many-body Schr\"odinger equation that describes the fully-heavy multiquark systems\footnote{Fully-heavy multiquark systems are going to be considered here because non-relativistic phenomenological Hamiltonians are naturally accepted to describe the dynamics of heavy hadrons; moreover,  we are pursuing general statements about the stability hierarchy of the different multiquark arrangements allowed by QCD and, since this has not done before, we consider heavy quark sectors the correct environment to begin with leaving light quark dynamics for future work.}. This approach allows us to reduce the uncertainty of the numerical calculation, accounts for multi-particle correlations in the physical observables, and generalizes the quark-clustering picture. The quark model we use~\cite{Semay:1994ht, Silvestre-Brac:1996myf} has a pairwise interaction which is the most general and accepted one: Coulomb$\,+\,$linear-confining$\,+\,$hyperfine spin-spin; therefore, our analysis should provide some rigorous statements about the mass location of the all-heavy multiquark ground states with different clustering assumptions. Note, too, that the model parameters were constrained by a simultaneous fit of $36$ mesons and $53$ baryons, with a range of agreement between theory and experiment around $10-20\%$, which can be taken as an estimation of our predictions shown here.


\noindent\emph{2.- Theoretical framework}.\,---\, Fully-heavy ground state systems can be described by the following Hamiltonian:
\begin{equation}
H = \sum_{i=1}^{\text{n-part.}}\left(m_i+\frac{\vec{p}^{\,2}_i}{2m_i}\right) - T_{\text{CM}} + \sum_{j>i=1}^{\text{n-part.}} V(\vec{r}_{ij}) \,,
\label{eq:Hamiltonian}
\end{equation}
where $m_{i}$ is the quark mass, $\vec{p}_i$ is the momentum of the quark, and $T_{\text{CM}}$ is the center-of-mass kinetic energy. Since chiral symmetry is explicitly broken in the heavy quark sector, the two-body potential, $V(\vec{r}_{ij})$, can be deduced from the one-gluon exchange and confining interactions; \emph{i.e.}
\begin{equation}
V(\vec{r}_{ij}) = V_{\text{OGE}}(\vec{r}_{ij}) + V_{\text{CON}}(\vec{r}_{ij})\,.
\end{equation}
The one-gluon exchange potential is given by
\begin{align}
V_{\text{OGE}}(\vec{r}_{ij}) &= \frac{1}{4} \alpha_{s} (\vec{\lambda}_{i}\cdot
\vec{\lambda}_{j}) \Bigg[\frac{1}{r_{ij}} \nonumber \\
&
- \frac{2\pi}{3m_{i}m_{j}} \delta^{(3)}(\vec{r}_{ij}) (\vec{\sigma}_{i}\cdot\vec{\sigma}_{j}) \Bigg] \,,
\end{align}
where $\alpha_s$ is the strong coupling constant, $\vec{\lambda}$ are the $SU(3)$-color Gell-Mann matrices, $\vec{\sigma}$ denote the Pauli spin matrices and the $\delta^{(3)}(\vec{r}_{ij})$ is replaced by a smeared function that reads as
\begin{equation}
\delta^{(3)}(\vec{r}_{ij}) \to \kappa \, \frac{e^{-r_{ij}^2/r_0^2}}{\pi^{3/2}r_{0}^3} \,,
\end{equation}
with $\kappa$ a quark model parameter, and $r_0 = A \left(\frac{2m_im_j}{m_i+m_j}\right)^B$ a regulator which depends on the reduced mass of the quark--(anti-)quark pair.

Lattice-QCD has demonstrated that multi-gluon exchanges produce an attractive linearly rising potential, which is proportional to the interquark distance~\cite{Bali:2005fu}. This is usually modeled as 
\begin{equation}
V_{\text{CON}}(\vec{r}_{ij}) = (b\, r_{ij} + \Delta) (\vec{\lambda}_{i}\cdot
\vec{\lambda}_{j}) \,,
\end{equation}
where $b$ is the confinement strength and $\Delta$ is a global constant fixing the origin of energies.

\begin{table}[!t]
\caption{\label{tab:parameters} Quark model parameters used herein and taken from AL1 potential in Refs.~\cite{Semay:1994ht, Silvestre-Brac:1996myf}.}
\begin{ruledtabular}
\begin{tabular}{llc}
Quark masses & $m_c$ (GeV) & 1.836 \\
\hline
OGE          & $\alpha_s$        & 0.3802 \\
             & $\kappa$          & 3.6711 \\
             & $A$ (GeV)$^{B-1}$ & 1.6553 \\
             & $B$               & 0.2204 \\
\hline
CON          & $b$ (GeV$^2$)  &  0.1653 \\
             & $\Delta$ (GeV) & -0.8321 \\
\end{tabular}
\end{ruledtabular}
\end{table}

Table~\ref{tab:parameters} shows the quark model parameters relevant for this work. Note here that we are using the so-called AL1 potential proposed by Silvestre-Brac and Semay in Ref.~\cite{Semay:1994ht}, and applied extensively to the baryon sector in Ref.~\cite{Silvestre-Brac:1996myf}.

The application of Quantum Monte Carlo (QMC) methods to hadron physics has been scarce, basically because most known hadrons consisted on bound states of just two and three quarks. However, many of the recently discovered XYZ particles are candidates to be 4-, 5- and even 6-quark bound or resonance states and thus QMC algorithms can become a competitive tool to shed some light into the spectroscopy and structure of multiquark systems. In fact, after the seminal works studying the fully-heavy tetraquark systems~\cite{Bai:2016int, Gordillo:2020sgc}, ourselves and other colleagues have been applying the same technique to other conventional and exotic hadron systems~\cite{Gordillo:2021bra, Ma:2022vqf, Gordillo:2022nnj, Ma:2023int, Mutuk:2023oyz, Gordillo:2023tnz, Alcaraz-Pelegrina:2022fsi}.

The central idea behind the Diffusion Monte Carlo method (DMC) is to write the Schr\"odinger equation for $n$-particles in imaginary time ($\hbar=c=1$):
\begin{equation}
-\frac{\partial \Psi_{\alpha'}(\bm{R},t)}{\partial t} = (H_{\alpha'\alpha}-E_s) \Psi_{\alpha}(\bm{R},t) \,,
\label{eq:Sch1}
\end{equation}
where $E_s$ is the usual energy shift used in DMC methods, $\bm{R}\equiv(\vec{r}_1,\ldots,\vec{r}_n)$ stands for the position of $n$ particles and $\alpha$ denotes each possible spin-color channel, with given quantum numbers, for the $n$-particles system. The function $\Psi_{\alpha}(\bm{R},t)$ can be expanded in terms of a complete set of the Hamiltonian's eigenfunctions as
\begin{equation}
\Psi_{\alpha}(\bm{R},t) = \sum_i c_{i,\alpha} \, e^{-(E_i-E_s)t} \, \Phi_{i,\alpha}(\bm{R}) \,,
\end{equation}
where the $E_i$ are the eigenvalues of the system's Hamiltonian operator. The ground state wave function, $\phi_{0,\alpha}(\bm{R})$, is obtained as the asymptotic solution of Eq.~\eqref{eq:Sch1} when $t\to \infty$, as long as there is overlap between $\Psi_{\alpha}(\bm{R},t=0)$ and $\phi_{0,\alpha}(\bm{R})$, for any $\alpha$-channel. This shall also provide us the ground-state mass of the different set of quarks given a particular set of quantum numbers $\alpha$.

From the paragraph above, one can deduce that the DMC method needs an initial approximation to the many-body wave function of the cluster, the so-called trial function, that should include all the information known \emph{a priori} about the hadron system. We chose the expression
\begin{align}
\Phi_{i,\alpha}(\bm{R}) &\equiv \Phi_i(\vec{r}_1,\ldots,\vec{r}_n;s_1,\ldots,s_n;c_1,\ldots,c_n) \nonumber \\
&
= \phi_i (\vec{r}_1,\ldots,\vec{r}_n) \nonumber \\ 
&
\times \big[ \chi_s (s_1,\ldots,s_n) \otimes \chi_c (c_1,\ldots,c_n) \big] \,,
\end{align}
where, explicitly, $\vec{r}_j$, $s_j$ and $c_j$ stand for the position, spin and color of the $j$-quark which is inside the $n$-quark cluster. 

In this work, we are going to consider hadron states that are eigenvectors of the angular momentum operator $L^2$ with eigenvalue equals to zero. This means that $\phi$ depends on the distance between pairs of quarks:
\begin{equation}
\phi (\vec{r}_1,\ldots,\vec{r}_n) = \prod_{j>i=1}^{n} \exp(-a_{ij} r_{ij}) \,.
\label{eq:radialwf}
\end{equation} 
Other alternatives to the radial part of the trial function are not considered in this work since, in principle, the DMC algorithm is able to correct its possible shortcomings and produce the exact masses of the arrangements~\cite{Gordillo:2020sgc}. Moreover, $a_{ij}$ are determined by the so-called cusp conditions, \emph{viz.} $a_{ij}$ are initially fixed in accordance to the boundary conditions of the problem.

The spin and color terms, $\chi_s$ and $\chi_c$, of the total wave function are written as linear combinations of the eigenvectors of the spin and color operators defined by:
\begin{equation}
F^2 = \left(\sum_{i=1}^{N_q} \frac{\lambda_i}{2} \right)^2\,, \quad\quad
S^2 = \left(\sum_{i=1}^{N_q} \frac{\sigma_i}{2}\right)^2.
\end{equation}
with eigenvalues $F^2=0$ (colorless functions) and $S=0$ or $1/2$, depending on whether the number of quarks in the multiquark system is even or odd, respectively. Those are the lowest possible eigenvalues for the spin operator and the only ones considered in this work. 

Since Eq.~\eqref{eq:radialwf} is symmetric with respect to the exchange of any two identical quarks, we have to produce spin-color combinations which are antisymmetric with respect to those exchanges in order to fulfill Pauli statistics. To do so, we apply the antisymmetry operator,
\begin{equation}
\mathcal{A} = \frac{1}{N} \sum_{{\alpha}=1}^N (-1)^P \mathcal{P_{\alpha}} \,,
\label{eq:antisymope}
\end{equation}
to the complete set of spin-color functions. In Eq.~\eqref{eq:antisymope}, $N$ is the number of possible permutations of the set of quark indexes, $P$ is the order of the permutation, and $\mathcal{P_{\alpha}}$ represents the matrices that define those permutations. Once constructed the matrix derived from the operator in Eq.~\ref{eq:antisymope}, we have to check if we can find any eigenvector with eigenvalue equal to one. If this is so, those combinations shall be the input of the DMC calculation.


\begin{table*}[!t]
\caption{\label{tab:results} Binding energies and Masses, in MeV, of the studied fully-charmed multiquark systems. We also provide relevant interquark mean-square radii, in fm$^2$. The subindexes in $\langle r_{ij}^2 \rangle$ represent $i$-quark and $j$-quark (or antiquark) within the $[1234]$-tetraquark, $[12345]$-pentaquark and $[123456]$-hexaquark.}
\begin{ruledtabular}
\begin{tabular}{lccccccc}
{\bf Tetraquarks} & Configuration & $E_B$ & $M$ & $\langle r_{12}^2 \rangle$ & $\langle r_{13}^2 \rangle$ & $\langle r_{14}^2 \rangle$ & $\langle r_{34}^2 \rangle$ \\
\hline
& $[(cc)(\bar c\bar c)]$ &  $-994$ & $6350$ & $0.26$ & $0.23$ & $\cdots$ & $0.26$ \\
& $[(c\bar c)(c\bar c)]$ & $-1312$ & $6032$ & $0.25$ & $0.49$ & $0.38$ & $0.25 $ \\
& $(c\bar c)-(c\bar c)$  & $-1334$ & $6010$ & $0.13$ & $\infty$ & $\cdots$ & $\cdots$ \\
\hline
\hline
{\bf Pentaquarks} & Configuration & $E_B$ & $M$ & $\langle r_{12}^2 \rangle$ & $\langle r_{14}^2 \rangle$ & $\langle r_{15}^2 \rangle$ & $\langle r_{45}^2 \rangle$ \\
\hline
& $[(cccc)\bar c]$ & $-986$ & $8194$ & $0.31$ & $\cdots$ & $0.31$ & $\cdots$ \\
& $[(cc)(cc)\bar c]$ & $-1159$ & $8021$ & $0.25$ & $0.27$ & $0.25$ & $0.26$ \\
& $[(ccc)(c\bar c)]$ & $-1257$ & $7923$ & $0.23$ & $0.61$ & $\cdots$ & $0.17$ \\
& $(ccc)-(c\bar c)$ & $-1281$ & $7899$ & $0.21$ & $\infty$ & $\cdots$ & $0.16$ \\
\hline
\hline
{\bf Open-charm} & Configuration & $E_B$ & $M$ & $\langle r_{12}^2 \rangle$ & $\langle r_{13}^2 \rangle$ & $\langle r_{15}^2 \rangle$ & $\langle r_{45}^2 \rangle$ \\
{\bf hexaquarks} & & & & & & & \\
\hline
& $[cccccc]$ & $-1114$ & $9902$ & $0.33$ & $\cdots$ & $\cdots$ & $\cdots$ \\
& $[(cc)(cc)(cc)]$ & $-1345$ & $9671$ & $0.25$ & $0.28$ & $0.28$ & $\cdots$ \\
& $[(ccc)(ccc)]$ & $-1400$ & $9616$ & $0.23$ & $\cdots$ & $0.65$ & $0.23$ \\
& $(ccc)-(ccc)$ & $-1420$ & $9596$ & $0.21$ & $\cdots$ & $\infty$ & $0.21$ \\
\hline
\hline
{\bf Hidden-charm} & Configuration & $E_B$ & $M$ & $\langle r_{12}^2 \rangle$ & $\langle r_{13}^2 \rangle$ & $\langle r_{15}^2 \rangle $ & $\langle r_{45}^2 \rangle$ \\
{\bf hexaquarks} & & & & & & & \\
\hline
& $[(ccc)(\bar c\bar c\bar c)]$ & $-1403$ & $9613$ & $0.23$ & $\cdots$ & $0.70$ & $0.23$ \\
& $(ccc)-(\bar c\bar c\bar c)$ & $-1420$ & $9596$ & $0.21$ & $\cdots$ & $\infty$ & $0.21$ \\
& $[(cc\bar c\bar c)(c\bar c)]$ & $-1624$ & $9392$ & $0.24$ & $0.22$ & $0.37$ & $0.15$ \\
& $(cc\bar c\bar c)-(c\bar c)$ & $-1661$ & $9355$ & $0.26$ & $0.23$ & $\infty$ & $0.13$ \\
\end{tabular}
\end{ruledtabular}
\end{table*}

\noindent\emph{3.- Results}.\,---\, Three multiquark structures: tetra-, penta- and hexa-quark systems, shall be under scrutiny in order to stablish the stable configurations for each exotic hadron.  We will concern ourselves only with arrangements made up of $c$ and $\bar{c}$ quarks. 

Let us then begin with the fully-charmed tetraquark system, $[cc\bar c\bar c] \equiv [(cc)(\bar c\bar c)]$; that is to say, we have two pairs of indistinguishable quarks and antiquarks, respectively. For this cluster, we have 2 color and 2 spin $S$=0 eigenfunctions, to be combined to give 2 totally antisymmetric color-spin functions. Those are then introduced in the DMC algorithm to get the binding energy, total mass and relevant mean-square radii shown in the first row of Table~\ref{tab:results}, section tetraquarks. We obtain a bound state in which the distances between all quarks are quite similar, indicating that it is a compact object. Our next step is to consider the $[(c\bar c)(c\bar c)]$ tetraquark, in which the $cc$ and $\bar{c}\bar{c}$  units are not considered to be undistinguishable, but they keep the tetraquark identity as a whole. In this case, we have the same 2 color and 2 spin eigenfunctions, but this time combined to produce 4 channels. The binding energy, total mass and mean-square radii produced by the DMC algorithm with those functions are shown in the second row of Table~\ref{tab:results}, section tetraquarks. Attending to the binding energy, this hadron is more bounded and then more stable than the former case; with respect to the interquark distances, one may conclude that mesons can be distinguished  clearly but the quarks in different mesons are further apart from each other.  In any case, we have a compact structure with finite distances between all quarks. The third row in Table~\ref{tab:results}, section tetraquarks, shows our results for the case in which the $a_{ij}$ coefficients in Eq. \ref{eq:radialwf} are different from zero only for the $c$-$\bar{c}$ in the same meson.  The color-spin functions are the same as in the previous $[(c\bar c)(c\bar c)]$ case.  The total mass is exactly the same as twice 
the meson mass in Ref.~\cite{Gordillo:2020sgc}, and the mean-square radii are also the same as for an isolated $c\bar{c}$. This situation is compatible with the infinite separation between quarks in different mesons. All this means that, even tough the ground state of a tetraquark corresponds to two mesons located infinitely apart, we can have two associated mesons close together forming a multiquark with a slightly larger mass. Above those, we will have a third possibility, a compact tetraquark. 

Concerning the fully-charmed pentaquark system, $[cccc\bar c]$, we have the following possible arrangements of quarks: $[(cccc)\bar c]$, $[(cc)(cc)\bar c]$, $[(ccc)(c\bar c)]$. In all cases, we have $3$ colorless color eigenfunctions to be combined with $5$ $S=1/2$ spin functions to produce $15$ color-spin functions.  For the first arrangement, in which we have four undistinguishable $c$ quarks, we found that there is only one possible antisymmetric combination of those $15$ wave functions if we consider all equal quarks undistinguishable. In the diquark-diquark-antiquark system, we have $4$ antisymmetric combinations that kept the antisymmetry of the quarks within each of the diquarks. In the baryon-meson cluster, we have $3$ combinations with the adequate symmetry. Their respective masses and quark-quark distances are in  Table~\ref{tab:results}, section pentaquarks. One can see that the $[(cccc)\bar c]$ configuration produces the highest energy with a clear compact structure among all involved quarks. This can be deduced from the equal values of the distances between particles. The following case, $(cc)(cc)\bar c$, has a mass lower than the first one in the spectrum and it is also a compact structure.  Then, we consider the $[(ccc)(c\bar c)]$ arrangement with a baryon-meson structure inside a pentaquark. In this case, the mass decreases again; moreover, the baryon$\,+\,$meson structure can be deduced from the separation between two quarks,  one in the baryon and another in the meson, $\sim 0.78\,\text{fm}$. Finally, the lowest energy state, and thus the more stable one, corresponds to the non-interacting baryon-meson system. Those results are the same as those obtained in Ref.~\cite{Gordillo:2020sgc}. It is worth mentioning that that $7899\,\text{MeV}$ mass listed in Table~\ref{tab:results} corresponds to the masses of a $ccc$ baryon with $S=3/2$ and a $J$/$\Psi$ meson due to the restrictions imposed by the $S=1/2$ of the whole pentaquark function. This reinforces the idea of having mesons and baryons as the most stable hadrons in nature.  However, as in the tetraquark case, we can, in principle, have a compact baryon-meson system slightly above the separated structure.  

Focusing now on the fully-charmed hexaquark system, we must consider two big families: the open-charm hexaquark, $[cccccc]$, and the hidden one, $[ccc\bar c\bar c\bar c]$. The number of their (color,spin)-eigenfunctions are $(5,5)$ and $(6,5)$, respectively. However, the final antisymmetric functions are just $1$, $2$ and $5$ for $[cccccc]$, $[(cc)(cc)(cc)]$ and $[(ccc)(ccc)]$ open-charm hexaquarks whereas, for hidden-charm hexaquarks, one finds $2$ antisymmetric functions for $[(ccc)(\bar c\bar c\bar c)]$ and $8$ for $[(cc\bar c\bar c)(c\bar c)]$. As for the open-charm hexaquark system, the possible quark configurations are shown in Table~\ref{tab:results}, section open-charm hexaquarks. One can see the pattern repeating itself, the binding energy becomes larger as we go from configuration $[cccccc]$ to configuration $[(cc)(cc)(cc)]$, passing through $[(ccc)(ccc)]$ and then $(ccc)-(ccc)$; therefore, the most stable situation is again having two $ccc$ non-interacting baryons. Besides, as shown by the interquark distances, configurations $[(cc)(cc)(cc)]$ and $[(ccc)(ccc)]$ are compact objects. As for the hidden-charm hexaquark family, it is remarkable to observe that the tetraquark$\,+\,$meson configuration has a larger binding energy than baryon-antibaryon case.  As all the cases studied before, either tetraquark-meson or baryon-antibaryon systems without interaction among them are more stable than their partner configurations within a compact hexaquark.  


\noindent\emph{4.- Summary}.\,---\, We have used a diffusion Monte Carlo method to solve a many-body Schr\"odinger equation that contains the most general and accepted pairwise Coulomb$\,+\,$linear-confining$\,+\,$hyperfine spin-spin interaction between heavy quarks and antiquarks.  We have found that for each colorless combination of $c$ and $\bar{c}$ quarks, we can have several compact clusters with different masses, that can, in principle, be detected separately. We also have shown that if the internal structure of the multiquark is made of pieces that are themselves individually colorless and with a definite value of the spin wave function, they separate to form smaller units.  


\begin{acknowledgments}
We acknowledge financial support from 
Ministerio Espa\~nol de Ciencia e Innovaci\'on under grant Nos. PID2020-113565GB-C22 and PID2022-140440NB-C22;
Junta de Andaluc\'ia under contract Nos. PAIDI FQM-205 and FQM-370.
The authors acknowledges, too, the use of the computer facilities of C3UPO at the Universidad Pablo de Olavide, de Sevilla.
\end{acknowledgments}


\bibliography{MultiquarkBreakingDMC}

\end{document}